# Improving Nonequilibrium Thermal Conduction in $CaF_2$ by Breaking the Phonon Bottleneck Effect

Mingde Qu[1,#], Minghuan Song[1,#], Yuxing Wang[1], Wudi Wang[1], Chenbo Zhang[1], Jiamin Quan[1], Jun Xu[1], Zhongwei Zhang[1,*], and Jie Chen[1,*]

[1]Center for Phononics and Thermal Energy Science, China-EU Joint Lab for Nanophononics, MOE Key Laboratory of Advanced Micro-structured Materials, School of Physics Science and Engineering, Tongji University, Shanghai 200092, People's Republic of China.

*Corresponding author: zhongwei@tongji.edu.cn(Z.Z.); jie@tongji.edu.cn(J.C.).

#These authors contributed equally to this work.

**ABSTRACT:** Heat dissipation in laser optical materials depends not only on their thermal conductivity but also on how absorbed energy is redistributed among various phonon modes. In calcium fluoride ($CaF_2$), rapid energy transfer from hot electrons to weakly propagating longitudinal optical (LO) phonons can create a nonequilibrium phonon bottleneck effect that limits efficient heat removal. Here, we propose selective phonon excitation as an effective strategy to break this bottleneck. Combining electron-phonon scattering calculations, phonon Boltzmann transport analysis, and nonequilibrium molecular dynamics simulations, we identify LO phonons as the primary modes that preferentially receive energy from hot carriers, leading to energy accumulation due to their inefficient transport and relaxation. Guided by three-phonon scattering phase-space analysis, selective excitation of transverse acoustic (TA) modes can accelerate LO relaxation and promote energy transfer toward heat-carrying acoustic modes. While sustained LO-phonon excitation reduces the thermal conductivity by approximately 12%, additional TA-mode excitation improves heat conduction by approximately 21% compared to the state limited by the LO-phonon bottleneck, surpassing the unexcited value. These findings demonstrate that controlling interbranch energy redistribution can counteract phonon-bottleneck-induced heat accumulation and reveal a working route for active thermal management in laser optical materials through selective control of phonon populations.

# I. INTRODUCTION

With the rapid development of high-power and high-energy laser technologies, heat dissipation has become a critical factor governing device efficiency, stability, and lifetime [1-3]. During laser operation, absorbed energy can be highly localized within microscopic volumes, producing severe thermal accumulation and steep temperature gradients near the irradiated region [2,4]. Such heating may induce thermal lensing, optical distortion, and laser damage [2,4]. Although external cooling techniques can remove heat at the device or system level, they do not directly regulate the ultrafast energy-transfer processes occurring inside laser-generated hotspots [5-9]. Improving heat dissipation at the microscopic level is therefore essential for the thermal management of high-energy laser systems [10-17].

Calcium fluoride ($CaF_2$) is widely used in optical components because of its broad transmission window and low refractive index [1,4]. Under irradiation conditions that generate hot carriers, energy transfer to the lattice can become strongly mode-selective. For instance, in polar crystals such as $CaF_2$, hot electrons can preferentially transfer energy to longitudinal optical (LO) phonons through Fröhlich coupling. These modes transport energy inefficiently because of their low group velocities. When the generation of LO phonons is faster than the relaxation, energy accumulates in the optical modes, creating a nonequilibrium phonon bottleneck [5-7]. Understanding this bottleneck therefore requires resolving both the selective generation of LO phonons and their subsequent energy transfer to heat-carrying modes.

Conventional approaches for engineering thermal transport include alloying,

isotope substitution, defect engineering, lattice symmetry, and the application of external fields [10-12,18,19]. For optical materials used in high-power laser systems, such as $CaF_2$, thermal management must preserve the essential properties for optical performance. Approaches that alter the crystal composition or introduce defects may compromise optical transparency or induce laser-damage resistance. On the other hand, mode-selective phonon excitation offers an alternative by regulating phonon populations without permanently modifying the crystal structure, potentially through appropriately tailored optical excitation. Recent studies [20,21] have demonstrated that targeted phonon excitation can modify heat conduction. However, the validity of this approach to mitigate the phonon bottleneck in dielectric materials used in laser systems remains unclear. In particular, a microscopic criterion for selecting auxiliary phonon modes and a clear connection between their effects on LO-phonon relaxation and macroscopic heat transport are still lacking.

In this work, we combine mode-resolved electron-phonon calculations, phonon Boltzmann transport analysis, and nonequilibrium molecular dynamics simulations with a validated machine-learning interatomic potential to investigate and mitigate the phonon bottleneck in $CaF_2$. We identify that the LO modes preferentially coupled to hot electrons, and further characterize their limited capability to transport the deposited energy. Three-phonon scattering phase-space analysis is then used to select auxiliary TA and TO modes that can facilitate LO-phonon relaxation, and their effects are assessed through occupation-dependent scattering calculations and mode-selective excitation simulations. We demonstrate that selective LO excitation slows thermal relaxation, whereas appropriate auxiliary-mode excitation promotes energy redistribution and increases the equivalent thermal conductivity.

## II. Model and Methods

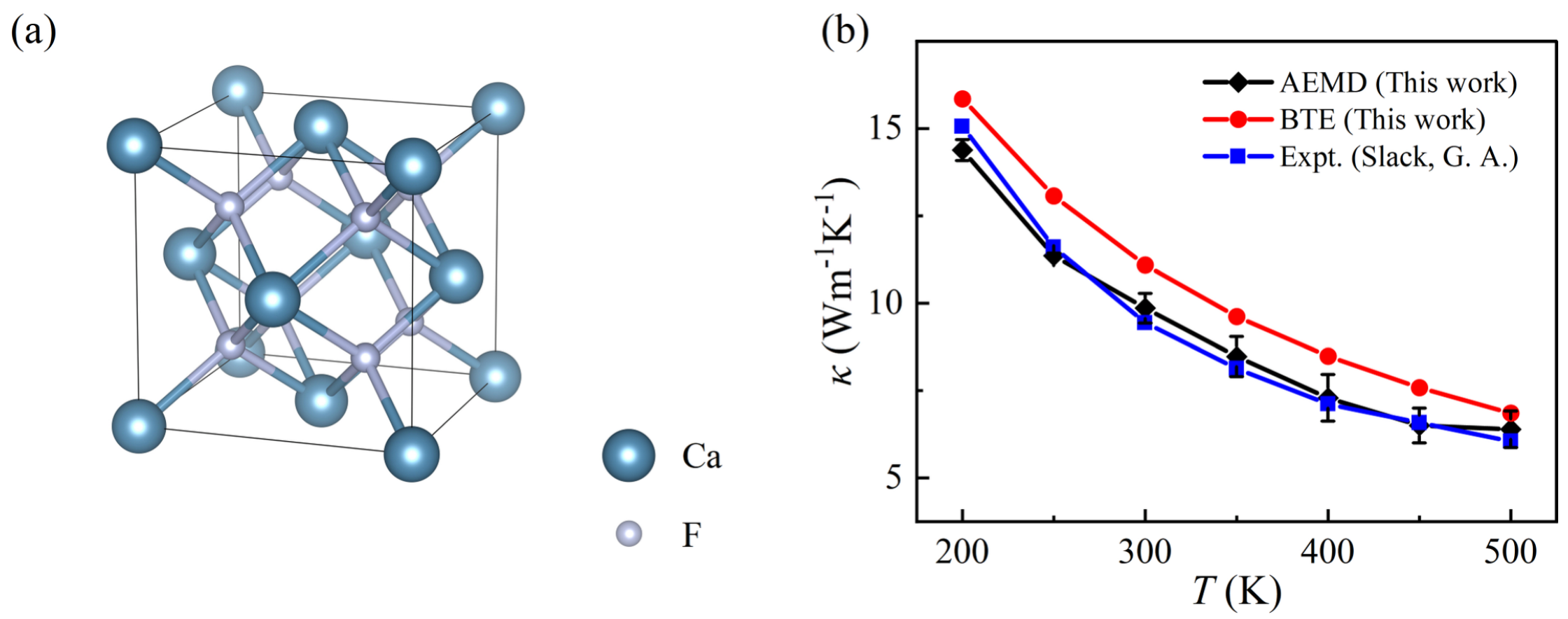

Figure 1. Schematic figure and thermal properties of $CaF_2$. (a) Fluorite crystal structure of $CaF_2$, with Ca and F atoms represented by heritage blue and lavender, respectively. (b) Temperature-dependent lattice thermal conductivity ($\kappa$) calculated in this work, compared with available experimental results [22].

### A. Machine learning potential and validation

$CaF_2$ crystal has the cubic fluorite structure with the space group $Fm\bar{3}m$. The conventional cubic cell, shown in Fig. 1(a), has a lattice constant of 5.463 Å and contains four Ca atoms and eight F atoms. This structure provides the basis for the atomistic calculations in this work. For the lattice-dynamics and Boltzmann transport calculations, a three-atom primitive cell containing one Ca atom and two F atoms is employed.

The atomic interactions in $CaF_2$ are described using a neuroevolution potential (NEP), trained following the protocol established in our previous works [23-25]. The detailed first-principles settings and dataset-generation procedure are provided in Ref. [23]. The NEP training is performed using the Graphics Processing Units Molecular

Dynamics (GPUMD) package [26-28]. The training is conducted for $10^5$ generations under the same training protocol as described in Ref. [23].

The results in Appendix confirm the convergence of the training loss functions, which decrease rapidly at the initial stage and gradually level off as training proceeds. The trained potential is evaluated by comparing its predicted energies and atomic forces with density-functional-theory (DFT) reference values for the test dataset. The root-mean-square errors (RMSEs) are 0.323 meV/atom for energies and 0.0294 eV/Å for atomic forces, indicating close agreement with the DFT results for the tested configurations.

To assess the validity for thermal calculations, we further compare the NEP and DFT predictions of phonon dispersion. As shown in Appendix, the NEP accurately reproduces both the acoustic and optical phonon branches across the Brillouin zone, which are consistent with literature results [29]. This agreement supports the accuracy of its description of the lattice dynamics of $CaF_2$ and provides a reliable basis for the subsequent thermal-transport simulations.

**B. Electron-phonon scattering**

Mode-resolved electron-phonon scattering calculations are performed in order to identify the phonon modes that are preferentially coupled to the photoexcited electronic population in $CaF_2$. The electronic structure is calculated within DFT using the Vienna Ab initio Simulation Package (VASP) [30]. The resulting electronic energies are used to characterize the electronic states relevant to the scattering analysis. The electron concentration is set to $n_{\mathrm{e}} \approx 6.7 \times 10^{18}\ \mathrm{cm}^{-3}$, and the electronic chemical potential is determined at an electronic temperature of 300 K to reproduce this concentration.

Electron-phonon scattering is calculated using density-functional perturbation theory (DFPT) in conjunction with Perturbo [31]. The electronic and phonon Brillouin zones are sampled using 9×9×9 $k$-point and $q$-point meshes, respectively. The electron-phonon matrix elements are evaluated using DFPT, and the energy-conservation delta functions are approximated using Lorentzian broadening.

The electron-phonon matrix element connecting the electronic states $(n, \boldsymbol{k})$ and $(m, \boldsymbol{k}+\boldsymbol{q})$ through phonon mode $\lambda = (\boldsymbol{q}, s)$ is defined as [32-34]:

$$g_{mn\lambda}(\boldsymbol{k}) = \langle \psi_{m,\boldsymbol{k}+\boldsymbol{q}} | \Delta_\lambda V_{\mathrm{KS}} | \psi_{n,\boldsymbol{k}} \rangle, \tag{1}$$

where $\psi_{n,\boldsymbol{k}}$ is the electronic Bloch state and $\Delta_\lambda V_{\mathrm{KS}}$ is the phonon-induced perturbation of the self-consistent potential. The imaginary part of the electron-phonon-induced phonon self-energy was given by [32,35]:

$$\mathrm{Im}\Pi_\lambda = \frac{\pi}{N_{\boldsymbol{k}}} \sum_{mn\boldsymbol{k}} |g_{mn\lambda}(\boldsymbol{k})|^2 \left(f_{n,\boldsymbol{k}} - f_{m,\boldsymbol{k}+\boldsymbol{q}}\right) \delta\left(\varepsilon_{m,\boldsymbol{k}+\boldsymbol{q}} - \varepsilon_{n,\boldsymbol{k}} - \hbar\omega_\lambda\right), \tag{2}$$

where $\varepsilon_{n,\boldsymbol{k}}$ and $f_{n,\boldsymbol{k}}$ are the electronic energy and occupation, respectively, and $N_{\boldsymbol{k}}$ is the number of sampled electronic wavevectors. Under the convention adopted here, $\mathrm{Im}\Pi_\lambda$ characterizes the relative electron-phonon scattering contribution associated with phonon mode $\lambda = (\boldsymbol{q}, s)$. To represent the hot-electron population relevant to the present study, the electronic states included in the mode-resolved analysis are restricted to the energy window from 10.70 to 11.30 eV.

### C. Phonon Boltzmann transport

The harmonic and anharmonic interatomic force constants (IFCs) of $CaF_2$ are obtained from first-principles calculations using a 6×6×6 supercell. The lattice thermal conductivity ($\kappa$) is obtained by solving the linearized phonon Boltzmann transport equation (BTE) [29,36-39] using the iterative scheme implemented in ShengBTE on a 12×12×12 $q$-point mesh [36,40].

Within the iterative solution of the linearized BTE, the lattice thermal conductivity tensor $\boldsymbol{\kappa}$ is expressed as [40,41]:

$$\kappa_{\alpha\beta} = \frac{1}{N_q V_0} \sum_{\lambda} C_\lambda v_{\lambda\alpha} F_{\lambda\beta} , \tag{3}$$

where $N_{\boldsymbol{q}}$ is the number of $q$-points in the Brillouin zone, $V$ is the volume of the primitive cell, $C_\lambda$ is the mode heat capacity, $v_{\lambda\alpha}$ is the group velocity of phonon mode $\lambda\ (\boldsymbol{q}, s)$ along the α direction, and $F_{\lambda\beta}$ is the $\beta$ component of the mode-resolved mean free displacement obtained from the iterative solution of the linearized BTE. The iterative BTE accounts for the coupling between different phonon modes through anharmonic three-phonon scattering processes [41,42]. The scattering rates are determined from the third-order IFCs by considering the corresponding energy- and crystal-momentum-conserving absorption and emission processes [40,41].

The mode-resolved $\kappa$ contributions are obtained from the corresponding terms in Eq. (3), and the spectral thermal conductivity $\kappa(\omega)$ is calculated by accumulating these contributions within discrete frequency intervals. The BTE calculations are further used to analyze the phonon group velocities, intrinsic lifetimes, three-phonon phase space, and occupation-dependent scattering properties.

### D. Three-phonon scattering and phase-space analysis

To identify auxiliary phonon modes that can facilitate the relaxation of the bottleneck LO modes, we analyze the three-phonon scattering phase space of $CaF_2$. For a target mode $\lambda$, absorption and emission processes involve pairs of modes $\lambda'$ and $\lambda''$ that satisfy energy and crystal-momentum conservation. For a target phonon mode $\lambda(\boldsymbol{q}, s)$, the kinematic three-phonon scattering phase spaces are defined as [41]:

$$P_{\lambda}^{\pm} = \frac{1}{N_{\boldsymbol{q}}} \sum_{\lambda',\lambda''} \delta_{\boldsymbol{q} \pm \boldsymbol{q}' - \boldsymbol{q}'', \boldsymbol{G}} \delta(\omega_{\lambda} \pm \omega_{\lambda'} - \omega_{\lambda''}), \tag{4}$$

where $\boldsymbol{G}$ is a reciprocal lattice vector, and the + and − signs denote the absorption and emission processes, respectively.

The phase space measures the availability of allowed scattering channels, whereas the scattering rates also depend on the third-order anharmonic interaction matrix elements. We therefore use phase-space analysis to screen candidate auxiliary modes and assess their effects on LO relaxation through phonon scattering calculations and molecular dynamics simulations.

To examine the effect of auxiliary-mode populations, we perform an occupation-perturbation analysis using ShengBTE. For the selected auxiliary modes, the equilibrium Bose–Einstein occupations ($n_{\lambda}^{BE}$) are scaled by a dimensionless factor $f$:

$$n_{\lambda} = f n_{\lambda}^{BE} = \frac{f}{\exp(\hbar\omega_{\lambda}/k_B T) - 1}. \tag{5}$$

Here, $f = 1$ gives the equilibrium occupation and $f > 1$ increases the selected-mode population for mode $\lambda$, while all other modes retain their equilibrium occupations. The modified occupations are substituted into the population factors of the three-phonon collision terms to recalculate the mode-dependent scattering rates and effective lifetimes. This analysis probes the sensitivity of phonon scattering to prescribed changes in auxiliary-mode populations; it does not solve the fully nonlinear nonequilibrium BTE self-consistently.

### E. Approach-to-equilibrium molecular dynamics

The nonequilibrium thermal response of $CaF_2$ is investigated using approach-to-equilibrium molecular dynamics (AEMD) [43-48], which tracks the relaxation of an initially imposed temperature difference. The simulation domain comprises 200 unit cells along the transport direction ($x$) and 20 unit cells along each of the two transverse directions, with periodic boundary conditions imposed in all three directions. Its two halves are maintained at 350 and 250 K using Langevin thermostats for $10^6$ steps. The thermostats are then removed, and the temperature evolution is monitored for another $10^6$ steps. A timestep of 0.5 fs is used throughout.

The temperature difference between the two regions, $\Delta T(t) = T_{hot}(t) - T_{cold}(t)$, is fitted to an exponential decay,

$$\Delta T(t) = \Delta T_0 \exp(-t/\tau), \tag{6}$$

where $\Delta T_0$ is the initial temperature difference and $\tau$ is the characteristic thermal relaxation time. Then, $\kappa$ is calculated as [43],

$$\kappa = \frac{C_V L^2}{4\pi^2 \tau}, \tag{7}$$

where $L$ is the cell length along the transport direction and $C_V$ is the volumetric heat capacity. The latter is obtained from equilibrium MD simulations as $C_V = (1/V)(\partial E/\partial T)$, using a fit to the total energy over a range of temperatures. Results are averaged over 5 independent trajectories for each simulation condition. The temperature-dependent $\kappa$ obtained from AEMD agrees well with the experimental measurements and our independent BTE calculations, as shown in Fig. 1(b). This agreement supports the accuracy of the interatomic potential and the AEMD procedure for describing thermal transport in unexcited $CaF_2$, providing a reliable reference for the subsequent mode-selective excitation simulations.

The transient nature of AEMD allows for examining how selective phonon excitation modifies the relaxation of an imposed temperature inhomogeneity. Conventional equilibrium MD methods evaluate transport coefficients from equilibrium fluctuations, whereas sustained mode-selective excitation maintains a nonequilibrium phonon population. In steady-state NEMD, the additional energy supplied by selective excitation must be explicitly accounted for in the heat balance and heat-flux analysis. Here, the AEMD-based protocol allows us to track the decay of the temperature difference under a prescribed excitation protocol and compare the thermal relaxation of different selectively excited states. Because energy continues to be supplied during this process, the fitted relaxation time reflects the combined effects of energy injection, redistribution, and transport. Under continuous mode-selective excitation, we define an equivalent $\kappa$ from the fitted decay time of the temperature difference using the same AEMD relation.

### F. Mode-selective phonon excitation

To establish controlled nonequilibrium phonon populations, mode-selective excitation is incorporated into the AEMD simulations by repeatedly rescaling selected modal velocity components during thermal relaxation. The modal velocity coordinate of mode $\lambda$ is obtained by projecting the atomic velocities onto its normalized eigenvector [21,49,50]:

$$\dot{Q}_\lambda = \sum_i \sqrt{m_i} \boldsymbol{e}_{i\lambda}^* \cdot \boldsymbol{v}_i \,, \tag{8}$$

where $m_i$, $\boldsymbol{v}_i$, and $\boldsymbol{e}_{i\lambda}$ denote the mass, velocity, and normalized polarization vector of atom $i$, respectively. To scale the kinetic energy of the selected mode by a factor $M$, the atomic velocities are updated as

$$\boldsymbol{v}_i' = \boldsymbol{v}_i + \left(\sqrt{M} - 1\right)\frac{\dot{Q}_\lambda \boldsymbol{e}_{i\lambda}}{\sqrt{m_i}}. \tag{9}$$

Here, $M = 1$ leaves the modal velocity unchanged, whereas $M > 1$ increases the modal kinetic energy. The excitation is applied within the first 10% of the simulation domain along the transport direction $x$.

To establish the LO-bottleneck state, selected high-frequency LO modes are excited to represent energy accumulation in weakly propagating optical modes. Auxiliary-excitation simulations additionally excite the TA or TO modes selectively through the three-phonon scattering phase-space analysis to examine their effects on LO relaxation. The injected energy is monitored, and the temperature evolution is analyzed using the AEMD procedure described above.

# III. RESULTS AND DISCUSSION

## A. Microscopic origin of the nonequilibrium phonon bottleneck

The formation of a nonequilibrium phonon bottleneck is governed by the competition between electronic energy deposition and subsequent phonon energy redistribution and transport, as illustrated in Fig. 2(a). We first examine the mode selectivity of electron–phonon coupling in $CaF_2$. Figure 2(b) presents the mode- and wavevector-resolved electron-phonon scattering contributions at an electron concentration of $n_\mathrm{e} \approx 6.7 \times 10^{18}\ \mathrm{cm}^{-3}$, considering electronic states within the energy window of 10.70–11.30 eV. The scattering contributions are concentrated on selected high-frequency optical branches near the Γ point, indicating that these modes preferentially receive energy from excited electrons.

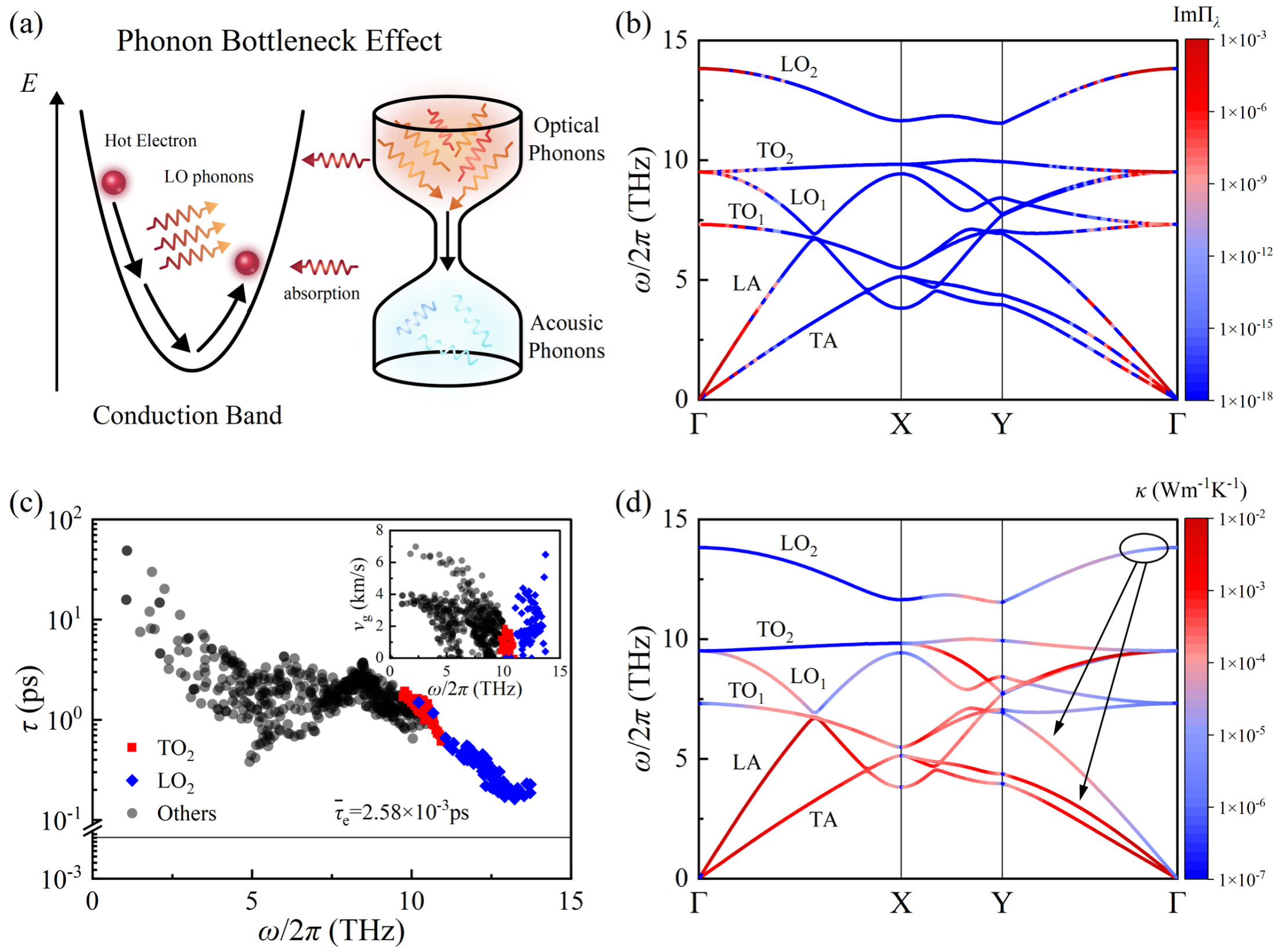

Figure 2. Electron-phonon coupling and phonon transport characteristics of the nonequilibrium phonon bottleneck in $CaF_2$. (a) Schematic illustration of the phonon bottleneck. Photoexcited hot electrons preferentially transfer energy to LO phonons through electron-phonon coupling. When LO-phonon generation outpaces anharmonic energy redistribution to acoustic modes, energy accumulates in the optical-phonon subsystem. (b) Phonon dispersion colored by the electron-phonon scattering contribution at an electron concentration of $n_e \approx 6.7 \times 10^{18}\ \mathrm{cm}^{-3}$, highlighting the optical modes strongly coupled to the electronic population. (c) Intrinsic phonon lifetimes as a function of frequency, with selected $LO_2$ and $TO_2$ modes highlighted. The horizontal line draws the electronic relaxation time $\bar{\tau}_e = 2.58 \times 10^{-3}$ ps for comparison. The inset shows the corresponding phonon group velocities. (d) Phonon dispersion colored by the mode-resolved contribution to the lattice thermal conductivity, showing the dominant contributions of low-frequency acoustic modes and the relatively small contributions of high-frequency optical modes.

To understand the resulting energy accumulation, Fig. 2(c) compares the intrinsic phonon lifetimes with the characteristic electron energy-relaxation time, $\bar{\tau}_e = 2.58 \times 10^{-3}$ ps. The electronic relaxation time is approximately two orders of magnitude shorter than the lifetimes of the selected high-frequency optical phonon modes, indicating that electronic energy can be deposited into these modes faster than it is redistributed through intrinsic phonon scattering. This timescale separation favors the buildup of a nonequilibrium optical-phonon population. Moreover, the selected $LO_2$ modes exhibit small group velocities [inset of Fig. 2(c)], limiting their ability to transport the deposited energy directly in the lattice system.

Figure 2(d) further reveals a mismatch between the modes receiving electronic energy and those responsible for heat conduction. While electronic energy is preferentially deposited into phonon modes near the Γ point, equilibrium lattice thermal transport is dominated by low-frequency LA and TA modes. The relatively small $\kappa$ contributions of $LO_2$ modes indicate that efficient heat removal requires energy redistribution toward the heat-carrying acoustic branches. Together, these results suggest that rapid electronic energy deposition, combined with limited phonon energy redistribution and weak direct transport, can lead to local energy accumulation and the formation of a nonequilibrium phonon bottleneck effect for high frequency $LO_2$ modes.

To investigate its consequences for thermal transport, we reproduce the sustained LO-phonon excitation background in molecular dynamics simulations through repeated mode-selective velocity rescaling, without explicitly evolving the electronic degrees of freedom. Hereafter, LO modes refer specifically to the highest-frequency $LO_2$ phonon branch. This approach enables us to quantify the impact of the phonon bottleneck on thermal relaxation and explore whether additional excitation of auxiliary phonon modes can improve heat transport.

### B. Breaking the phonon bottleneck through selective phonon excitation

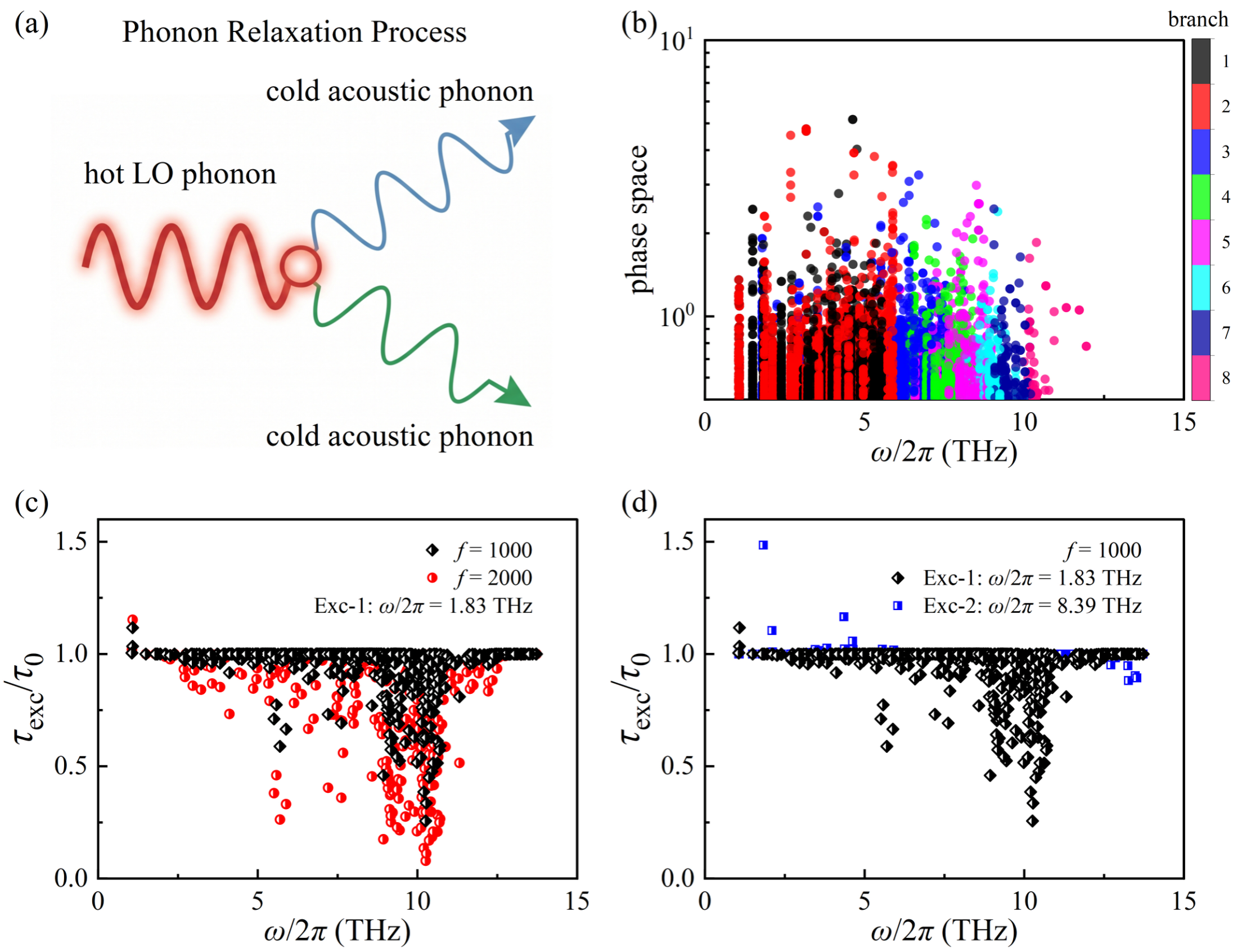

Figure 3. Breaking the phonon bottleneck through selective excitation of auxiliary phonon modes. (a) Schematic of three-phonon decay processes involving a target hot LO phonon. (b) Branch-resolved three-phonon phase space involving the selected LO modes across the Brillouin zone. (c) Ratio of phonon lifetimes under auxiliary-mode excitation to their equilibrium values as a function of the Bose–Einstein occupation scaling factor $f$ for a fixed auxiliary mode. Here, $f = 1$ corresponds to the equilibrium occupation. (d) Phonon lifetime ratios obtained by exciting auxiliary modes of different frequencies and branch characters at the same occupation scaling factor $f$. The TA mode at 1.83 THz and the TO mode at 8.39 THz are labeled Exc-1 and Exc-2, respectively.

After identifying the high-frequency LO modes as the primary recipients of electronic energy, we next investigate which auxiliary phonon modes can facilitate their relaxation and thereby break the phonon bottleneck effect. We focus on three-phonon emission processes in which a target LO phonon decays into two lower-energy phonons, subject to energy and crystal-momentum conservation, as illustrated in Fig. 3(a). Following Eq. (4), we fix the target LO mode while varying the other two phonon modes over the sampled phonon states. For each auxiliary mode, we sum over all allowed partner modes in the three-phonon scattering process to evaluate its contribution to the decay phase space of the target LO phonon. By resolving this contribution with respect to the branch and wavevector of the auxiliary mode, we identify the specific phonon modes that can participate in the LO-decay channels.

Figure 3(b) reveals a strongly nonuniform distribution of the decay phase space, with pronounced maxima for selected TA modes at 1.83 THz. These maxima indicate favorable energy and momentum matching with the targeted LO modes, providing a criterion for selecting auxiliary modes that may facilitate LO relaxation.

To evaluate the influence of auxiliary-mode excitation on LO relaxation, we perform an occupation-perturbation analysis using ShengBTE. The equilibrium Bose–Einstein occupations of the selected auxiliary modes are scaled by a dimensionless factor $f$, such that $n_{\lambda'} = f n_{\lambda'}^{\mathrm{BE}}$, while all other modes retain their equilibrium occupations. Thus, $f = 1$ corresponds to the equilibrium population, whereas $f > 1$ represents an enhanced auxiliary-mode population. The modified occupations enter the three-phonon scattering terms, allowing us to evaluate the resulting changes in scattering rates and effective phonon lifetimes. This analysis isolates the scattering response to prescribed phonon populations without explicitly evolving the nonequilibrium distribution.

Figure 3(c) shows the normalized LO-phonon lifetime as a function of the occupation scaling factor for the selected TA mode at 1.83 THz. Increasing the TA-mode population progressively shortens the effective lifetimes of the bottleneck LO modes, indicating enhanced LO-involving scattering. Figure 3(d) further compares the responses to different auxiliary modes. At the same occupation scaling factor, excitation of the TA mode at 1.83 THz produces a larger change in the LO and acoustic phonon lifetimes than excitation of the TO mode at 8.39 THz.

These results demonstrate that the effectiveness of auxiliary-mode excitation depends strongly on phonon branch and frequency. The selected TA modes provide favorable LO-decay channels and exhibit a stronger occupation-dependent scattering response, suggesting a potential route to accelerate LO-energy redistribution. We next examine whether this microscopic scattering mechanism translates into improved thermal transport using molecular dynamics simulations.

### C. Suppression of thermal conductivity under the phonon bottleneck

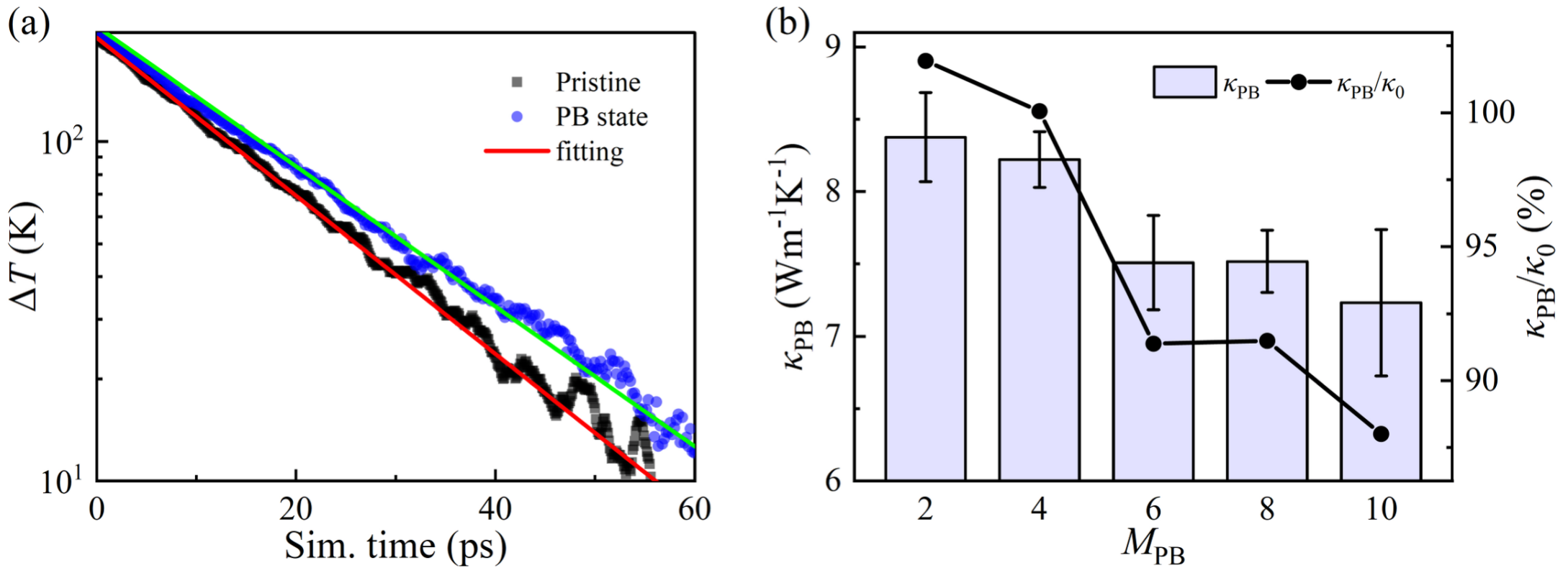

Figure 4. Formation of the phonon bottleneck under selective LO-mode excitation. (a) Time evolution of the temperature difference $\Delta T(t)$ in the unexcited reference state and the phonon bottleneck (i.e., PB) state with a modal kinetic-energy scaling factor of $M_{\mathrm{PB}} = 4$. Symbols represent the AEMD simulation results, and solid lines are fits to Eq. (6). (b) Equivalent thermal conductivity of the phonon-bottleneck state, $\kappa_{\mathrm{PB}}$, as a function of $M_{\mathrm{PB}}$. The right axis gives the normalized conductivity, $\kappa_{\mathrm{PB}}/\kappa_0$, where $\kappa_0$ is the thermal conductivity of the unexcited reference state.

To investigate the impact of the LO-phonon bottleneck on thermal transport, we selectively excite the highest-frequency LO modes within 10.21–10.22 THz, which exhibit strong electron–phonon coupling but limited direct heat-carrying capability, as identified in Fig. 2. Following the procedure described in Sec. II.F, the kinetic energies of these modes are repeatedly rescaled within the first 10% of the simulation domain along the transport direction. The excitation strength is governed by the modal kinetic-energy scaling factor $M_{\mathrm{PB}}$. Here, $M_{\mathrm{PB}} = 1$ represents the unexcited reference state, whereas $M_{\mathrm{PB}} > 1$ repeatedly injects energy into the selected modes to model the phonon bottleneck (PB) effect. This procedure establishes a sustained nonequilibrium LO-phonon population during the thermal relaxation process.

Figure 4(a) compares the decay of temperature-difference between the unexcited and LO-excited states at $M_{\mathrm{PB}} = 4$. The slower decay under LO excitation indicates that the imposed temperature imbalance persists longer, reflecting suppressed thermal relaxation. By fitting the AEMD temperature decay using Eq. (6), we obtain the characteristic relaxation times and the corresponding thermal conductivities shown in Fig. 4(b).

$\kappa$ decreases monotonically with increasing LO-excitation strength, from 8.22 $\mathrm{Wm^{-1}K^{-1}}$ in the unexcited state to 7.23 $\mathrm{Wm^{-1}K^{-1}}$ at $M_{\mathrm{PB}} = 4$, corresponding to a reduction of approximately 12%. This suppression reflects the sustained population of weakly propagating LO modes, which limits the efficiency of energy redistribution toward heat-carrying acoustic modes and consequently slows down thermal relaxation.

### D. Improved Thermal conductivity by breaking the phonon bottleneck

The preceding results show that sustained energy deposition into weakly propagating LO modes impedes thermal relaxation. Breaking this bottleneck requires promoting energy redistribution from the LO reservoir toward heat-carrying acoustic modes. The phase-space analysis and occupation-dependent lifetime calculations in Fig. 3 identify selective phonon excitation as a possible route: increasing the populations of modes that participate in allowed three-phonon scattering processes involving LO phonon can accelerate LO relaxation. We therefore examine whether this scattering response leads to improved thermal transport in the nonequilibrium MD simulations.

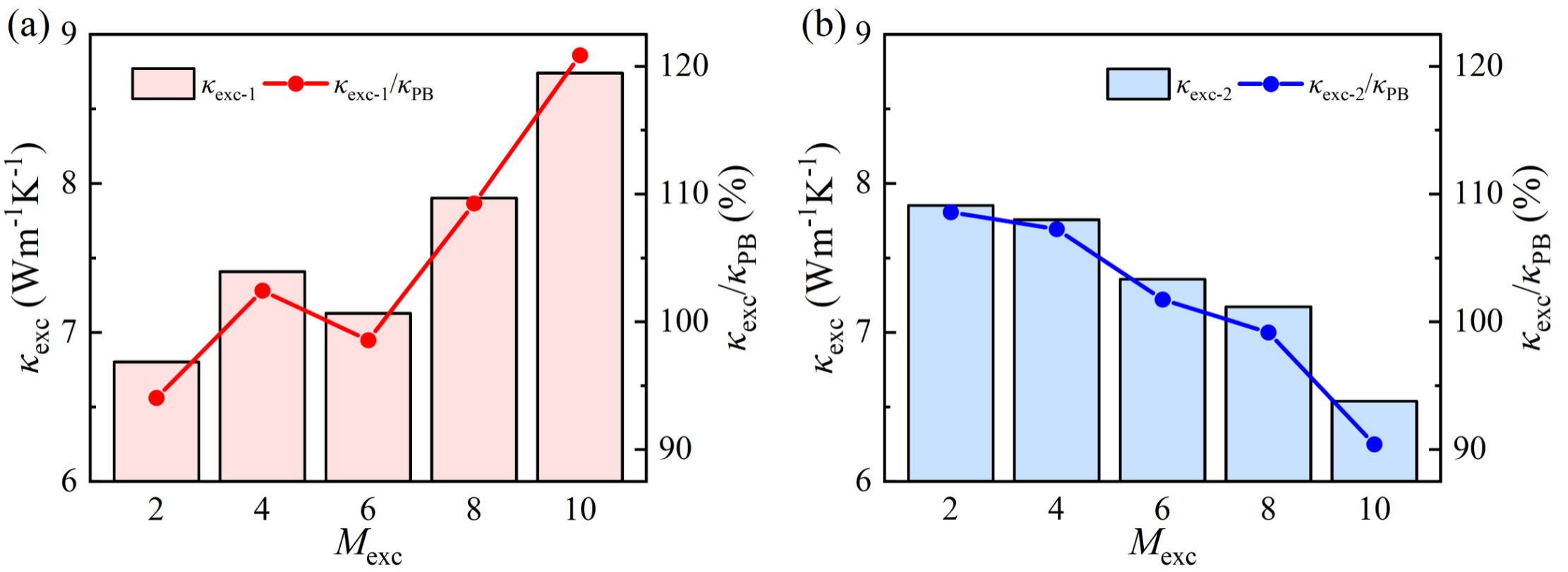

Figure 5. Breaking the phonon bottleneck through selective phonon excitation. Thermal conductivity under phonon excitation ($\kappa_{\text{exc}}$) as a function of the modal kinetic-energy scaling factor for excitation (exc) of (a) the TA mode at 1.83 THz ($M_{\text{exc}-1}$) and (b) the TO mode at 8.39 THz ($M_{\text{exc}-2}$). The LO-mode excitation strength is held constant to maintain the phonon-bottleneck condition. Results are averaged over eight independent MD simulations.

Based on the resolved phase space analysis, a TA mode at 1.83 THz is selected as the primary auxiliary mode. A TO mode at 8.39 THz is also included for comparison. In all simulations shown in Fig. 5, the LO-bottleneck excitation is fixed at $M_{\text{PB}} = 10$. The modal kinetic-energy scaling factor of the auxiliary mode, $M_{\text{exc}}$, is then varied while keeping the LO excitation fixed. The cases $M_{\text{exc}-1} = 1$ and $M_{\text{exc}-2} = 1$ both correspond to the reference LO-bottleneck state without additional auxiliary-mode excitation.

Figure 5(a) shows a nonmonotonic dependence of $\kappa_{\text{exc}}$ on $M_{\text{exc}-1}$ under TA-

mode excitation. Starting from 7.23 $Wm^{-1}K^{-1}$ in the reference bottleneck state, $\kappa_{\mathrm{exc}}$ initially decreases at low excitation strength and then increases monotonically with further increases in $M_{\mathrm{exc-1}}$, reaching 8.74 W $Wm^{-1}K^{-1}$ at $M_{\mathrm{exc-1}} = 10$. This increase represents an enhancement of approximately 21% relative to the bottleneck state, and also exceeds the unexcited reference value of 8.22 $Wm^{-1}K^{-1}$. Selective TA-mode excitation thus counteracts the suppression of thermal relaxation associated with LO-energy accumulation, even while the LO excitation is maintained. In contrast, excitation of the TO mode at 8.39 THz leads to a progressive decrease in $\kappa_{exc}$ with increasing excitation strength ($M_{\mathrm{exc-2}}$), as shown in Fig. 5(b). This mode dependence is consistent with the results in Fig. 3, where excitation of the selected TA mode produces a much weaker impact in the effective LO lifetimes than excitation of the TO mode.

These findings reveal a microscopic mechanism in which selective phonon excitation promotes energy redistribution from the LO reservoir toward heat-carrying acoustic modes, thereby breaking the phonon bottleneck and improving heat dissipation. Experimentally, laser-induced transverse acoustic phonons and terahertz acoustic parametric excitation have been reported in other material systems [50,51], suggesting possible routes toward populating the selected TA modes. Realizing this approach in $CaF_2$ will require an excitation scheme compatible with the frequency, wavevector, and polarization of the targeted mode.

## IV. Conclusion

In summary, we have identified the microscopic origin of the nonequilibrium phonon bottleneck effect in $CaF_2$ and demonstrated a working route to overcome it through selective phonon excitation. This bottleneck arises from a mismatch between rapid electronic energy deposition into selected LO phonon modes and slower energy redistribution and spatial transport in the lattice system. Selective excitation of TA modes participating in favorable three-phonon scattering processes promotes LO-energy relaxation and facilitates energy transfer toward heat-carrying acoustic modes.

Under sustained LO excitation, the equivalent thermal conductivity decreases by approximately 12%. Excitation of the selected TA mode reverses this suppression, resulting in an enhanced equivalent thermal conductivity by approximately 21% relative to the bottleneck state and also above the unexcited reference value. These findings highlight the role of mode-dependent energy redistribution in nonequilibrium heat transport and establish selective phonon excitation as a potential strategy for improving heat dissipation in laser optical materials.

**ACKNOWLEDGMENTS**

This project is supported in part by grants from the National Natural Science Foundation of China (Grants No. 12675046, No. 12475037 and No. 52632001), the Science and Technology Commission of Shanghai Municipality (Grant No. 24520711200), the Basic Research Program of Jiangsu (Grant No. BK20244002), and the Fundamental Research Funds for the Central Universities (Grants No. 22120260019 and No. 22120260066). J.C. is supported by the Shanghai Leading Talent Program of Eastern Talent Plan (Grant No. LJ2025020).

**DATA AVAILABILITY**

The data that support the findings of this article are available from the authors upon reasonable request.

# Appendix

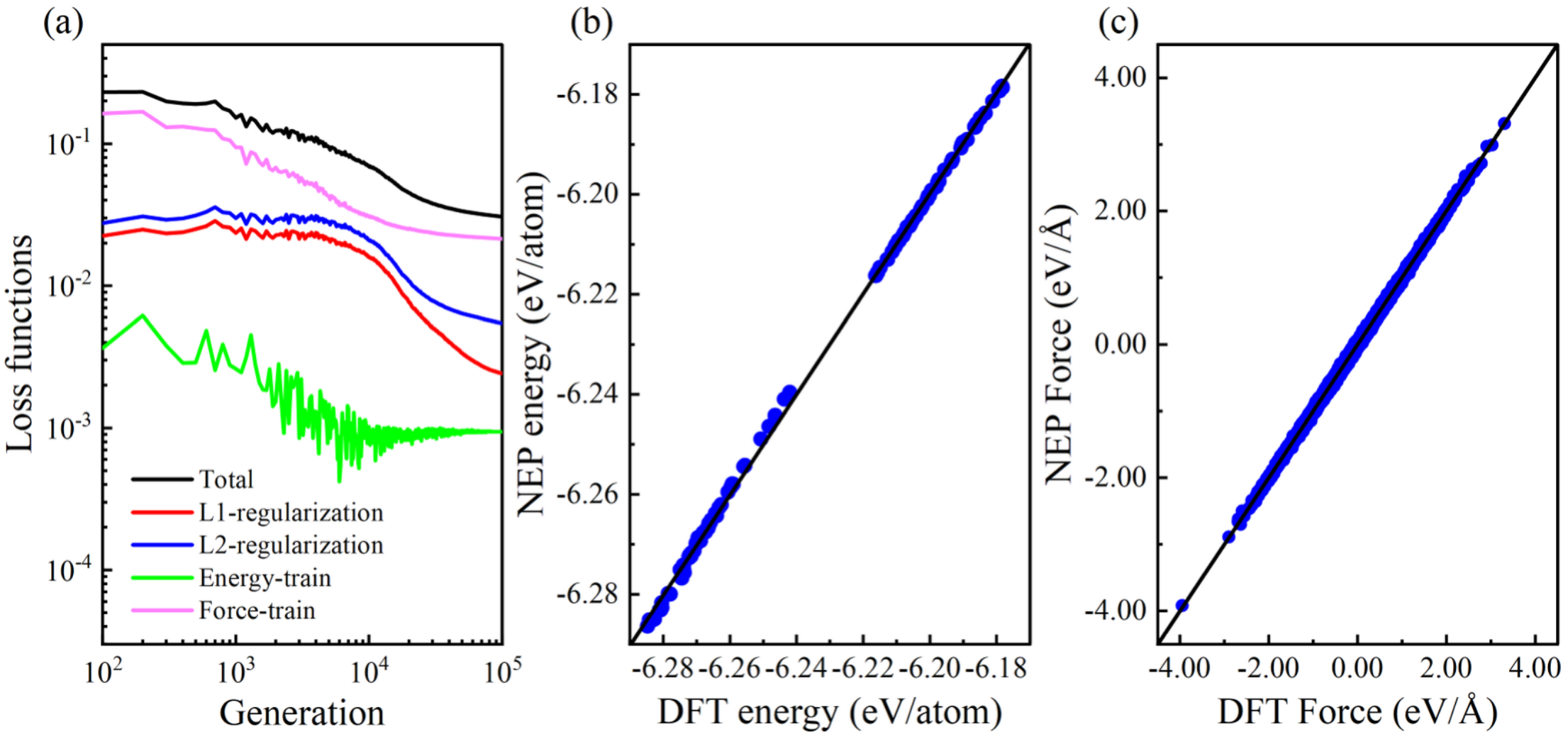

Figure 6. Training convergence and prediction accuracy of the neuroevolution potential (NEP). (a) Evolution of the L1 and L2 regularization terms and the energy, force, and virial training losses with generation. (b, c) Comparison of NEP-predicted energies and atomic force components with the corresponding density-functional-theory (DFT) reference values for the test dataset. The diagonal lines indicate perfect agreement between the NEP predictions and DFT references.

Figure 6 in the Appendix summarizes the training convergence and prediction accuracy of NEP. Fig. 6(a) shows the evolution of the L1 and L2 regularization terms and the training losses associated with energies, atomic forces, and virials. The energy and force losses decrease substantially during training and approach plateaus at later generations, indicating convergence of these training objectives. The predictive accuracy of the trained potential is further evaluated using the test dataset, as shown in Figs. 6(b) and 6(c), which compare the NEP-predicted energies and atomic force components with the corresponding DFT reference values. The data points lie close to the diagonal lines, with root-mean-square errors (RMSEs) of 0.323 meV/atom for energies and 0.0294 eV/Å for atomic forces. These results demonstrate close agreement with the DFT references for the tested configurations.

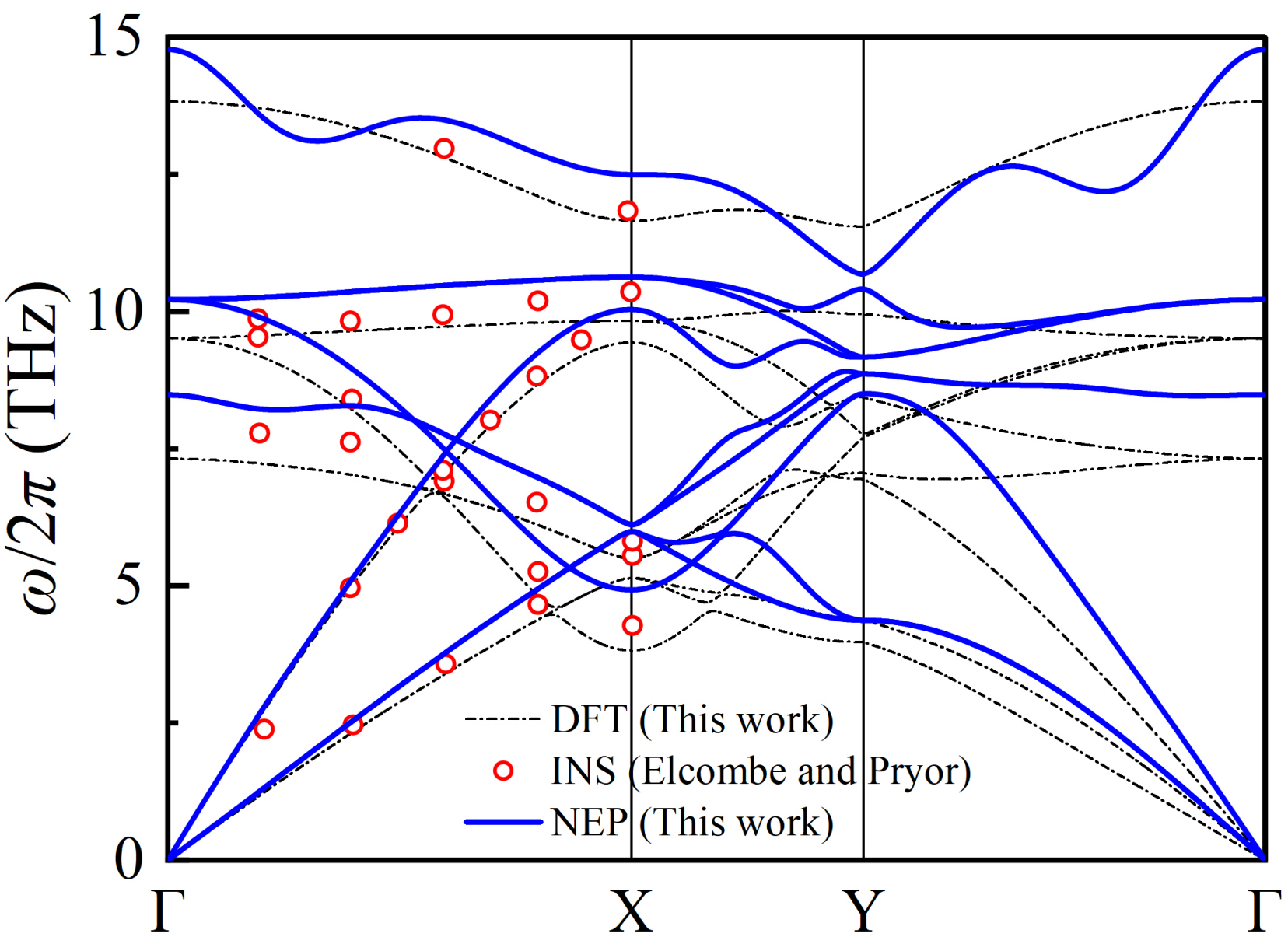

Figure 7. Comparison of the phonon dispersions of $CaF_2$ calculated using DFT (black dash-dotted lines) and NEP (blue solid lines) with the experimental inelastic neutron-scattering data of Elcombe and Pryor (red open circles) [29].